\documentclass[%
 reprint,
 showpacs,
 showkeys,
 preprintnumbers,
 amsmath,amssymb,
 aps,
 prl,
  longbibliography,
 ]{revtex4-1}

\usepackage[breaklinks=true,colorlinks=true,anchorcolor=blue,citecolor=blue,filecolor=blue,menucolor=blue,pagecolor=blue,urlcolor=blue,linkcolor=blue]{hyperref}
\usepackage{graphicx}
\usepackage{xcolor}

\begin{document}

\title{What Maxwell's demon could do for you}

\author{Karl Svozil}
\affiliation{Institute of Theoretical Physics, Vienna
    University of Technology, Wiedner Hauptstra\ss e 8-10/136, A-1040
    Vienna, Austria}
\email{svozil@tuwien.ac.at} \homepage[]{http://tph.tuwien.ac.at/~svozil}

\date{\today}

\begin{abstract}
In this highly speculative Letter it is argued that, under certain physical conditions, Maxwell's demon might be capable of breaking the second law of thermodynamics, thereby allowing a perpetual motion machine of the second kind, by accessing single particle capabilities.
\end{abstract}

\pacs{05.20.-y, 05.70.-a, 05.20.Gg}
\keywords{Maxwell's demon, ensemble theory, second law of thermodynamics}
\maketitle

Nearly at the end of his treatise on the {\em Theory of Heat}~\cite[pp.~338-339]{Maxwell-1871},  James Clerk Maxwell noted that,
by utilizing the molecular constitution of bodies, the production of a temperature gradient without expenditure of work cannot be outrightly excluded
even in a totally isolated system (associated with a microcanonical ensemble).
Thereby, he introduced an agent (later termed {\em Maxwell's intelligent demon}~\cite{thomson-1874})
sorting individual gas constituents according to their velocities into different regions of space.

Subsequent discussions might have suffered from the presumption to validate the second law of thermodynamics~\cite{Earman1998435,Earman19991},
thereby in various ways  ``explaining'' why Maxwell's demon must fail~\cite{maxwell-demon2}.
The present line of ``exorcism of the demon'' is via the physical limits information and computation is subjected to~\cite{landauer-94}.
In particular, a complete cycle would require the demon to ``forget'' the information
it has been forced to accept in order to perform a selection, thereby effectively undergoing an irreversible
many-to-one evolution.
Any such irreversible process is associated with an energy dissipation which (allegedly) should be at least as high as the energy gained
through it separation procedure~\cite{penrose-oliver-70,bennett-82}. With these premises, the combined system, including the demon, cannot
only produce a temperature gradient, because any such separation of constituents according to their energy is at least counterbalanced
by the subsequent necessity of ``unthinking;'' that is, resetting the demon's mind.

Alas, as history shows, too often the appearance of prudence serves as a disguise for orthodoxy:
after all, it is possible to transmute mercury into gold~\cite{PhysRev.60.473}, sailing faster than the wind, or
actively eavesdropping on quantum cryptographic protocols certified to be ``unconditionally secure''
by compromising the classical non-quantum  as well as the quantum communication channels~\cite{benn-84}.
A slight change of the rules, capabilities or assumptions may result in huge changes of capacities.

One of the assumptions entering the presently accepted exorcism is
based on the fact that the actions of the demon, in particular also its internal functions and processes, as compared to the rest of the (supposedly isolated) system,
are equal in magnitude.
In terms of statistical physics, the phase space volume
occupied by the demon $\Omega_d  $, and the  phase space volume
occupied by the rest of the system $\Omega_r  $, should be of the same order; that is,
$
O(\Omega_d )=O(\Omega_r )
$ for all times;
their relative magnitude slightly fluctuating around one for all times of a complete cycle.
The combined system, including the demon subsystem and the rest, occupies a region of phase space
equivalent to the product of the single phase space regions
$\Omega   = \Omega_d \Omega_s $.
Since entropy should be an extensive (additive) quantity,  Boltzmann's equation
(with the number of microstates identified with the phase space volume in accord with the Boltzmann hypothesis)
suggests a logarithmic relation between phase space volumes and entropies; i.e.,
$S   = k_B {\rm log} (\Omega_d  \Omega_r )
=  k_B {\rm log} \Omega_d  + k_B {\rm log} \Omega_r
= S_d   + S_r  $, where $k_B \approx 10^{-23} {\rm m}^2 {\rm kg} / ({\rm s}^2 {\rm K})$ stands for Boltzmann's constant.

Contemporaray exorcism argues that every sorting action of the demon,
associated with a decrease of entropy of the rest system $\Delta S_r$,
is compensated by an increase $\Delta S_d$ of the demon's ``mind'' memory which is at least as big as the entropy decrease caused by the demon; that is,
$\Delta S_d + \Delta S_r \ge 0$.
This is often demonstrated~\cite[pp.~927-929]{bennett-82} by a modified argument utilizing a one-molecule Szil\'ard engine~\cite[pp.~843--844]{Szilard-1929}
whose phase space volume is effectively reduced by a factor of two, resulting in $\Delta S_r = - k_B {\rm log} 2$.
In order to complete the cycle, one bit of the demon's memory has to be resetted; an irreversible, two-to-one transition causing an increase
of entropy at least as big as  $\Delta S_d =  k_B {\rm log} 2$ \cite{landauer:61}.

Alas, this argument fails if the reduction of phase space of the remaining system is much bigger than the increase in phase space of the demon; that is,
for
$
-\Delta \Omega_r \gg \Delta \Omega_d
$;
or, stated in terms of entropies, for
$
-\Delta S_r \gg \Delta S_d
$.
Whether this can be excluded remains unknown.
Indeed,
one may conceive of a system whose constituents are ``huge'' with respect to the demon.
Suppose, for example, that the system consists of Billard balls and a very tiny demon controlling the entry to a hole.

Likewise one may consider a ``relay'' demon operating with a ``weak interaction,''
whereas the rest of the system is dominated by a strong force.
In such a system, the demon might be able to produce a thermal separation and thus an energy gradient convertible into work,
thereby outpacing the energy required for the restoration of its ignorance.
The resulting combined system would effectively constitute a perpetual motion machine of the second kind
(see Ref.~\cite{PhysRevE.57.3846} for a different model).
It may even not be entirely unreasonable to imagine a system which may be able to propel a vehicle by extracting heat from the surroundings.
This could be effectively conceptualized by a modified Szil\'ard engine with an additional final stage, acquiring the same amount of heat
which was previously transformed to work from a thermal heat bath.

Another challenge against the second law of thermodynamics arises from the possibility to ``compress'' the information messing up the demon's memory.
Because the erasure of a maximally compressed (in terms of algorithmic information theory~\cite{chaitin3,calude:02,li-vitanyi-2008}) information,
the associated phase space volume, instead of $\Omega_d$,
would require merely the erasure of the algorithmic information content (sometimes referred to as Kolmogorov complexity)
$H(\Omega_d)\le \Omega_d$ thereof, it would effectively render a similar scenario as discussed earlier.

This issue has been pronounced solved~\cite{zurek,li-vitanyi-2008,681318}.
The optimal efficiency could only be reached by requiring the demon to store information without redundancy;
that is in its highest possible compression.
Note that, due to  information theoretic incompleteness, that is, due to
quantitative bounds on determining the algorithmic information~\cite[Section~8.2]{calude:02},
the highest possible compression for all practical purposes~\cite{bell-a} is unattainable, nonoperational
and in general incomputable.
In this view, the demon would need hypercomputing capacities to do a good job (albeit never breaching the second law of thermodynamics);
all else would result in too many erased bits,
thereby greatly reducing the gain obtained by a temperature (velocity) sort.
One of the issues is the assumption that, instead of
the raw phase space volume $\Omega_d$, its  maximal compression $\Omega_d^\ast$
is inserted for the calculation of the entropy, resulting in a modified entropy definition
$S^\ast   = k_B [ H(\Omega_d){\rm log} 2  +  {\rm log} \Omega_r ]$,
which is then subjected to the second law of thermodynamics~\cite[Example and claim~8.6.1]{li-vitanyi-2008}.

Quite generally, neither Maxwell's original scenario nor Szilard's engine seem to suggest any crucial dependence on compression.
Indeed, it may not be necessary and conceptually impossible for the demon's operation to assume that the operations result in a highly compressed memory --
after all, Szil\'ard's engine involves only a single particle (or bit); any such state cannot be further compressed without complete loss of information.

For the sake of the argument, suppose the ensemble is chosen
such that the demon's memory fills up with some subsequence of length $L\gg 1$ of
(a) Chaitin's random halting probability~\cite{rtx100200236p}, or
(b) Sierpinski's computable~\cite{Becher2002947} absolutely normal number, or
(b) Champernowne's constant normal in base~2~\cite{bailey-crandall-2001}
    which is constructed of all the integers concatenated in lexicographical order, or any
(c) other simply normal~\cite{Hertling:jucs_8_2:simply_normal_numbers_to} (i.e. unbiased) number such as $0.\dot{1}\dot{0}$.
If the demon performed his duties correctly, the amount of separation in all cases would be maximal, and the resulting work extracted at temperature $T$ could be
$ST= k_B T L{\rm log} 2$.
This is supposed to be counterbalanced by resetting the demon's memory, in which finite sequences of random (for the halting probability) and
compressible bits are stored.
Even if the operation of the demon and the rest of the system is assumed to be on an equal footing (which need not be the case, as mentioned earlier)
because the demon is capable of compression, only the random case (a) would satisfy the second law of thermodynamics.
Actually, any regularity in the ensemble could be used to violate it,
and to extract more work than is required for resetting the demon's mind.
How can one accept the notion that the second law is saved only by the assumption of absolute randomness?
Because the more regular $\Omega_r$ appears, the more work can be extracted from it.
In particular, any periodic sorting such as the one in case (c) requires very little algorithmic information on the demon's side,
and could readily be used to extract work.

To conclude this Letter, let me summarize the two main issues raised against present exorcisms.
First, the actions of and inside the demon's mind might be regarded small with respect to the rest of the system.
Second, compressing the demon's mind, despite all assurances to the contrary, might reduce the dissipative costs of erasing the information acquired.
For these reasons the author would agree with Earman and Norton~\cite{Earman19991} that ``the demon lives.''


%

\end{document}